\documentclass[a4paper,twocolumn,american,prl]{revtex4}
\usepackage[T1]{fontenc}
\usepackage[latin9]{inputenc}
\usepackage{graphicx}

\makeatletter

\usepackage{graphicx}
\usepackage{babel}
\makeatother
\begin{document}
\title{Electronic properties of linear carbon chains: resolving the controversy}

\author{Amaal Al-Backri$^{1,2
}$, Viktor Z\'olyomi$^{1}$, and Colin J. Lambert$^{1}$}
\affiliation{
$^{1}$Physics Department, Lancaster University, LA1 4YB, Lancaster,
United Kingdom}

\affiliation{
$^{2}$Baghdad University, College of Science,Al-Jaderyia campus, Baghdad, Iraq}


\begin{abstract}
Literature values for the energy gap of long one-dimensional carbon chains vary from as little as 0.2 eV to more than 4 eV.
To resolve this discrepancy, we use the GW many-body approach to calculate the band gap $E_g$ of an infinite carbon chain.
We also compute the energy dependence of the attenuation coefficient $\beta$ governing the decay with chain length of the
electrical conductance of long chains and compare this with recent experimental measurements of the single-molecule
conductance of end-capped carbon chains. For long chains, we find $E_g = 2.16$ eV and an upper bound for $\beta$ of  $0.21$ \AA$^{-1}$.
\end{abstract}
\maketitle

One of the possible routes to sustaining the development of sub-10nm computational architectures is the
use of single molecules as electronic components \cite{1}. A variety of methods have been developed for measuring  the conductance of single molecules in
contact with metallic electrodes \cite{2,3,4,5,6a,6b,7,8} and the availability of these techniques has stimulated many investigations into  molecular structure-property relationships \cite{11a,11b,12,13,x15,14a,14b,15}.
For the purpose of connecting molecular-scale devices, it is desirable to identify single-molecule wires, whose electrical conductance G decays only slowly with molecular length L and for this reason much interest has focused on the length dependence of the molecular conductance of homologous
series of pi-conjugated oligomers \cite{13,16,18,19,20a,20b,21,22,23,24,y3,y4,y5,y6,y7}. In almost all molecular junctions measured to date, the Fermi energy $E_F$ of the metallic electrodes lies in the HOMO-LUMO gap of the molecule and therefore electron transport occurs via tunnelling. Consequently for large enough L, the electrical conductance varies as $G = A exp{-\beta(E_F)L}$, where the pre-factor A depends on details of the electrode-molecule interface, whereas for large L, the attenuation coefficient $\beta(E_F)$ is a property of the molecular wire and the Fermi energy $E_F$ of the electrodes. Since $\beta(E_F)$ is minimised by reducing the offset between $E_F$ and either the HOMO or LUMO levels, it is desirable to minimise the HOMO-LUMO gap and maximise the electronic coupling between adjacent monomer segments of long wires. Since the rotation of such segments can break conjugation and reduce the pi coupling between segments \cite{13,14a,14b}, it is advantageous to introduce constraints,  which prevent rotation or alternatively to use wires made from segments which have no rotational degrees of freedom \cite{x15}.

In this article, we investigate the electronic properties of the thinnest and toughest example of the latter,
formed from one-dimensional carbon chains. Such conjugated polyyne molecules \cite{x1} are formed from a linear chain of
sp-hybridized carbon atoms with alternating single and triple bonds  and delocalized orbitals along the carbon backbone.
Polyynes and shorter sp oligomers (oligoynes) have been proposed for a variety molecular electronics applications \cite{x8}.
As the length L of the sp-chain increases, polyynes become increasingly unstable and therefore longer chains with up to 44
carbons are invariably stabilised by terminating them with bulky aryl or organometallic
groups \cite{x25,x26,eisler2005polyynes,paalsson2010efficient}. Alternative strategies involve synthesising them inside
carbon nanotubes \cite{XCNT} or identifying such chains linked to graphene electrodes \cite{xxx}. Optical spectroscopic
studies in solution confirm that these systems exist as alternating single and triple bonds with a low HOMO-LUMO gap.
Theoretical studies have predicted that polyynes should have excellent intramolecular electron-and charge-transport properties
and recent experimental studies have probed the length dependence of the conductance of oligoynes with up to 8 carbon
atoms. \cite{x19,x32}.
Despite these extensive investigations, the electronic properties of polyynes remain controversial. On the one hand,
literature values for the extrapolated energy gaps $E_g$ of infinitely-long polyynes vary widely from 0.3 eV to over
4 eV (see Figures S.1 and S.2 of the Supplementary Material (SM) for a summary of theoretical and experimental literature
values)\cite{suppl}. On the other hand, recently-measured $\beta$ values for oligoynes range from 0.006 \AA$^{-1}$ to 0.32 \AA$^{-1}$
depending on the end caps and measurement environment. Clearly this situation is unsatisfactory. If the properties of such
simple carbon chains cannot be accurately determined, then what hope is there of accurately describing more complex molecular
wires? From an experimental viewpoint, these differences may arise from different experimental set ups and measurement
environments. From a theoretical viewpoint, the known inadequacies of density functional theory (DFT) in computing
HOMO-LUMO gaps are a major issue.

Here we aim to resolve this controversy through a careful ab initio theoretical calculation, based on the GW many-body approach, which goes beyond DFT.
We use first principles calculations to obtain the band gap of an infinite polyyne and compare it with existing
experimental data. We also compare semilocal with hybrid DFT and demonstrate the
inadequacy of the former to the study of Peierls distorted systems. We then use the GW approximation to
calculate the correct quasiparticle band gap. Finally, we use the GW band structure to parametrize
a tight-binding model and calculate the $\beta$ value of polyynes as a function of energy within
the band gap, thereby providing an upper bound for the value of $\beta$ in the large $L$ limit.

\begin{figure}
\includegraphics[width=0.9\columnwidth]{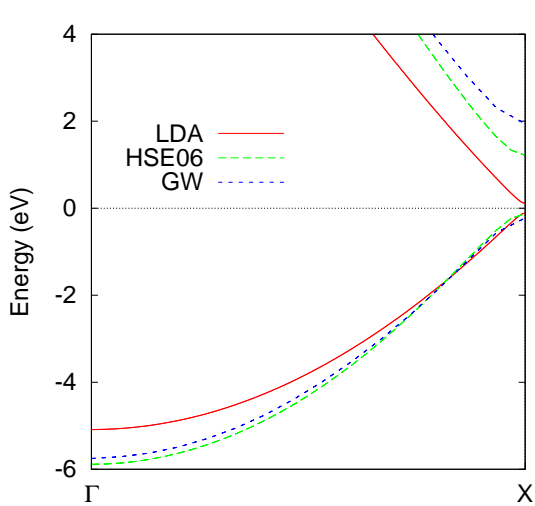}
\caption{
(Color online) Band structure of the infinite polyyne chain according to the GW approximation and density functional theory.
}
\label{Fig1_bandstruct}
\end{figure}

To demonstrate the limitations of DFT, we began by performing DFT calculations on an infinite polyyne using the plane-wave based
Vienna ab initio Simulation Package \cite{KresseG_1993_1,KresseG_1996_1}. We employed the Heyd-Scuseria-Ernzerhof 2006 (HSE06) hybrid
functional to optimize the atomic structure of polyyne and compared the results to earlier calculations using the local
density approximation (LDA) \cite{PhysRevB.72.155420}.
In the case of LDA, the lattice constant is 2.548 \AA~ and the bond length alternation (BLA) is 0.034 \AA,
while the HSE06 functional yielded a lattice constant of 2.554 \AA~ and a significantly larger BLA
of 0.090 \AA. The LDA result is not surprising as semilocal DFT is known to significantly
underestimate the BLA of Peierls-distorted systems \cite{SunG_2002_1,YangS_2004_1}. Note that neither the use of a localized
basis set nor the inclusion of gradient corrections in the generalized gradient approximation can resolve this problem.
Indeed we have confirmed using the SIESTA \cite{SolerJM_SIESTA_2002} code with a double $\zeta$ polarized basis that SIESTA and
VASP give the same result in the case of the LDA, and that the GGA and LDA results hardly differ. Therefore it is reassuring that the
hybrid HSE06 functional can treat Peierls distorted chains better. It is
important to note that the correct size of the BLA is crucial in order to obtain the correct band gap, as we discuss below.

We have calculated the band structure of an infinite polyyne in the optimal geometry using both the LDA and the HSE06 functional. The results
are compared in Figure \ref{Fig1_bandstruct}, which shows that the LDA band gap is vastly underestimated, being a mere 0.3 eV. In
contrast the HSE06 functional predicts a significantly larger band gap of 1.34 eV. The large difference between the two methods comes
from the superior description of both electronic excitations and Peierls distortion by the hybrid functional. However, even hybrid
density functionals can underestimate the band gap and therefore we calculated the many-body corrections to the band structure using the
GW approximation (see SM). We found that the GW band gap takes the significantly-larger value of 2.16 eV, which is in excellent agreement with experiments, as shown in Figure S.2 of the SM.
As shown in Figure 1, the shape of the valence band in the HSE06 functional and the GW approximation are almost the same, while the LDA differs
significantly from them.

While the GW approximation offers a powerful means of calculating the band gap, it is essential to initiate the calculation using the correct atomic
structure and BLA. This is well illustrated by recently published calculations, where the GW
approximation was applied on top of an LDA calculation which crudely underestimates the BLA  \cite{BadGWArxiv}.
This approach leads to a band gap of merely 0.41 eV despite the GW corrections. We have calculated the single-shot GW band gap in the LDA geometry as well, and we have found that in this case the band gap is merely 0.99 eV, as expected. The difference between
this and our value of 2.16 eV is a clear reminder that semilocal density functional theory
cannot be relied upon to predict the atomic structure of Peierls distorted systems.

\begin{figure}
\includegraphics[width=0.9\columnwidth]{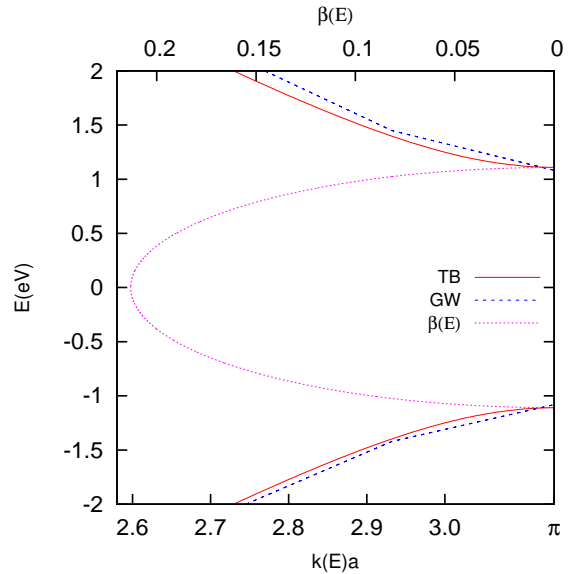}
\caption{
(Color online) The $\beta$ value (\AA$^{-1}$) of polyyne according to our tight-binding model (see text) as a function of energy (eV) within the band gap.
The tight-binding (TB) band structure around the gap is also plotted. The parameters of the tight-binding model were obtained
by fitting the TB band structure to the GW results, which are also shown for comparison.
}
\label{Fig2_beta}
\end{figure}

Now we turn our attention to the energy dependence of the attenuation coefficient $\beta$, which in the case of an infinite polyyne, can be identified with twice the imaginary part of the wave vector within the energy gap. To derive an analytical
formula for $\beta(E)$ as a function of the energy $E$, we introduce a tight-binding model, with one orbital per site, whose parameters are determined by fitting to the GW band structure of Figure 2.  The tight-binding lattice is shown in Figure S5 of the SM and contains nearest-neighbour hopping parameters $\gamma$ and $\alpha$, corresponding to long and short bonds respectively. As shown in the SM, this yields an energy gap of $E_g = 2\vert \gamma - \alpha \vert$ and the following formula for the $\beta$ value as a function
of energy:

\begin{equation}
\beta = -(2/a)ln\left(\vert \Theta (E)\vert - \sqrt{\Theta^{2}(E)-1}\right),
\end{equation}

\noindent where $\Theta = (E^2-(\alpha^2+\gamma^2))/2\alpha\gamma$. This expression holds for for energies within the energy gap, for which $\vert \Theta (E)\vert > 1$.
For energies within the band, the parameters $\alpha$ and $\gamma$  can be fitted to our first principles calculations and since the formula assumes electron-hole
symmetry we fitted the tight-binding band structure to the valence band of our GW calculations. The resulting energy dependence of $\beta(E)$ is shown in Figure \ref{Fig2_beta}. If $\beta_{expmt}$ is a measured experimental value, then this formula can
be used to calculate the Fermi energy$E_F$ by setting $\beta(E_F) = \beta_{expmt}$. Alternatively, if $E_F$ is known, then $\beta(E_F)$ is the predicted attenuation coefficient in the large L limit. As shown in Figure 2, the maximum value
of $\beta = 0.21$\AA$^{-1}$ is attained in the middle of the band gap.

Finally, let us compare our results for $E_g$ and $\beta$ with existing literature values.
The vast majority of the available experimental data is for short oligoynes with various end groups. These contain either the same end group
at opposite ends (symmetric arrangement) or different end groups (asymmetric arrangement). To compare with the above results for $\beta$ and $E_g$, we are interested in extrapolating these experimental values to the limit of infinite lengths.
Figures S.1 and S.2 of the SM compare existing theoretical and experimental literature values, respectively.
Clearly the theoretical results of Figure S.1 extrapolate to values in a quite broad range depending on the side
groups and the computational  method employed. This illustrates that DFT and  even hybrid DFT are unreliable for the quantitative study of polyynes. On the other hand, Figure S.2 shows that most of the experimental data  converge to the
vicinity of 2--2.3 eV, which is in excellent agreement with
our GW result of 2.16 eV.
We also note that in a previous study of polyynes synthesized inside multi-walled carbon nanotubes, resonance Raman
spectroscopy estimated the band gap at around 2.5 eV \cite{XCNT}. This is also consistent with
our result as the encapsulated polyyne had a finite length, hence its HOMO-LUMO gap should be larger than that of an infinite polyyne.

Comparison between the $\beta$ value of Figure 2 and existing experiments is less straightforward, because measurements are only
available on relatively short oligoynes with up to 8 carbon atoms. Futhermore the end caps on these short oligoynes can move charge
on and off the chains and therefore shift the energy gap relative to the Fermi energy of the electrodes. In addition, experiments are performed in a variety of ambient conditions and
 it is known that fluctuations due to for example the presence of
an environment can decrease the value of $\beta$ for electrons with energies lying in an energy gap \cite{cjl1} and increase $\beta$ for those lying outside the gap \cite{cjl2}.   Nevertheless for long enough
chains, Figure 2 shows that whatever the value of $E_F$, $\beta$ should lie in the range $ 0 < \beta < 0.21 \AA^{-1}$. Since the energy
gap of finite oligoynes is larger than the energy gap of an infinite polyyne, the maximum $\beta$ value of short chains is likely to be
larger than that of the infinite chain. Therefore this range of $\beta$ values is consistent with the experimenally measured range of
0.006 \AA$^{-1}$ to 0.32 \AA$^{-1}$.

In summary, we have presented in Figures 1 and 2 of the SM a summary of literature results,  which demonstrates that the precise values of the energy gaps of polyynes are rather controversial. To resolve this controversy, we have used the many-body GW approach to compute the band gap $E_g$ and attenuation coefficient $\beta(E)$ of an infinite polyyne. We have shown that semilocal density functional theory significantly underestimates $E_g$  and that
a correct description of the electronic structure of this material requires both a correct lattice structure, which can
be obtained via the hybrid HSE06 functional and the application of the GW quasiparticle corrections. We also provide an
analytical formula for the attenuation coefficient $\beta(E)$ by fitting to the GW band structure, which provides an upper bound for
the attenuation coefficient of long polyynes.

Details of the parameters used in the DFT and GW calculations, and the derivation of the tight-binding model can be
found in the Supplementary Material.

Work supported by the Marie Curie ERG project Carbotron (PERG08-GA-2010-276805), the Marie-Curie ITN MOLESCO, the Iraqi government, the EPSRC and OTKA K81492.
We thank Jen\H{o} K\"urti for valuable discussions.



\end{document}